\documentclass{aa}

\usepackage{graphicx}

\begin{document}

\title{Cepheid distances from infrared long-baseline interferometry}
\subtitle{II. Calibration of the Period--Radius and Period--Luminosity relations}
\titlerunning{Calibration of the P--R and P--L relations of Cepheids}
\authorrunning{P. Kervella et al.}

\author{
P. Kervella\inst{1,2},
D. Bersier\inst{3},
D. Mourard\inst{4},
N. Nardetto\inst{4}
\and
V. Coud\'e du Foresto\inst{1}
}

\offprints{P. Kervella}

\institute{LESIA, Observatoire de
Paris-Meudon, 5, place Jules Janssen, F-92195 Meudon Cedex,
France \and European Southern Observatory, Alonso de Cordova 3107,
Casilla 19001, Vitacura, Santiago 19, Chile
\and Space Telescope Science Institute, 3700 San Martin Drive, Baltimore, MD 21218, USA
\and Observatoire de la C\^ote d'Azur, D\'ept GEMINI, UMR6203, BP
4229, 06304 Nice Cedex 4, France }

\mail{pkervell@eso.org}

\date{Received ; Accepted}

\abstract{ Using our interferometric angular diameter measurements
of seven classical Cepheids reported in Kervella et
al.\,(\cite{kervella04}, Paper I), complemented by previously
existing measurements, we derive new calibrations of the Cepheids
Period--Radius (P--R) and Period--Luminosity (P--L) relations. We
obtain a P--R relation of $\log R = [0.767 \pm 0.009]\,\log P +
[1.091 \pm 0.011]$, only 1\,$\sigma$ away from the relation
obtained by Gieren et al.~(\cite{gieren98}). We therefore confirm
their P--R relation at a level of $\Delta(\log R) = \pm 0.02$.
We also derive an original calibration of the P--L relation, assuming
the slopes derived by Gieren et al.~(\cite{gieren98}) from LMC
Cepheids, $\alpha_K = -3.267 \pm 0.042$ and $\alpha_V = -2.769 \pm
0.073$.  With a P--L relation of the form $M_{\lambda} =
\alpha_{\lambda}\,(\log P - 1) + \beta_{\lambda}$, we obtain $\log P =
1$ reference points of $\beta_K = -5.904 \pm 0.063$ and $\beta_V =
-4.209 \pm 0.075$.  Our calibration in the $V$ band is statistically
identical to the geometrical result of Lanoix et
al.~(\cite{lanoix99}).
 }

\maketitle

%
\section{Introduction}
The Period--Luminosity (P--L) relation of the Cepheids is the basis of
the extragalactic distance scale, but its calibration is still
uncertain at a $\Delta M = \pm 0.10$\,mag level. Moreover, it is not
excluded that a significant bias of the same order of magnitude
affects our current calibration of this relation.  On the other hand,
the Period--Radius relation (P--R) is an important constraint to the
Cepheid models (see e.g. Alibert et al.~\cite{alibert99}).

Traditionally, there has been two ways to calibrate the P--L relation.
For Cepheids in clusters one can use main sequence fitting, assuming
that the main sequence is similar to that of the Pleiades. This method
has been questioned however, following the release of Hipparcos data
(e.g. Pinsonneault et al.~\cite{pinsonneault98}, but see also Pan et al.~\cite{pan04}
and Robichon et al.~\cite{robichon99}).
Another route to the P--L relation is the
Baade-Wesselink (BW) method where one combines photometry and radial
velocity data to obtain the distance and radius of a Cepheid.  Recent
applications of the BW method to individual stars can be found for
instance in Taylor et al.~(\cite{taylor97}) and Taylor \&
Booth~(\cite{taylor98}), while the calibration of the P--R and P--L
relations using BW distances and radii is demonstrated in Gieren,
Fouqu\'e \& Gomez~(\cite{gieren98}, hereafter GFG98).  A requirement
of this method is a very accurate measurement of the Cepheid's
effective temperature at all observed phases, in order to determine
the angular diameter. Interferometry allows us to bypass this step and
its associated uncertainties by measuring \emph{directly} the
variation of angular diameter during the pulsation cycle.  As shown by
Kervella et al.~(\cite{kervella04}, hereafter Paper~I) and Lane et
al.~(\cite{lane02}), the latest generation of long baseline visible
and infrared interferometers have the potential to provide precise
distances to Cepheids up to about 1\,kpc, using the interferometric BW
method (see Sect.~\ref{sect_context}).

The main goal of the present paper is to explore the application of
this technique to the calibration of the P--R and P--L relations, and
to verify that no large bias is present in the previously published
calibrations of these important relations.  Our sample is currently
too limited to allow a robust determination of the P-L relation,
defined as $M_\lambda = \alpha_\lambda (\log P - 1) + \beta_\lambda$,
that would include both the slope $\alpha_\lambda$ and the $\log P =1$
reference point $\beta_\lambda$.  However, if we suppose that the
slope is known {\it a priori} from the literature, we can derive a
precise calibration of $\beta_\lambda$.  In Sect.~\ref{P_R_sect}, we
present our determination of the P--R relation using new angular
diameter values from Paper~I, as well as previously published
interferometric and trigonometric parallax measurements.
Sect.~\ref{P_L_sect} is dedicated to the calibration of the P--L
relation reference points $\beta_\lambda$ in the $K$ and $V$ bands.
The consequences on the LMC distance are briefly discussed in
Sect.~\ref{lmc_dist}.

\section{Cepheid distances by interferometry \label{sect_context}}

We have obtained angular diameter measurements for seven Cepheids
with the VLT interferometer (Kervella et al. 2003, Paper 1). These
$K$-band measurements were made with the VINCI instrument
(Kervella et al.~\cite{kervella03}) fed by two 0.35m
siderostats. Several baselines were used, ranging from 60\,m to
140\,m. Our measurements, described in detail in Paper~I, have a
typical precision of 1 to 3 \%. This is good enough to actually
\emph{resolve} the pulsation of several Cepheids; in other words
we can follow the change in angular diameter. We have combined
these measurements with radial velocity data and derived a radius
and distance for four Cepheids of our sample. For the remaining three
stars, we were able to derive their mean angular diameters, but the pulsation
remained below our detection threshold. This sample was completed
by previously published measurements obtained with other instruments.

In the present work, we have retained the limb darkened (LD)
angular diameters $\theta_{\rm LD}$ provided by each author.
Marengo et al.~(\cite{marengo02}, \cite{marengo03})
have shown that the LD properties of Cepheids can be different
from those of stable stars, in particular at visible wavelengths. For the
measurements obtained using the GI2T (Mourard et al.~\cite{mourard97})
and NPOI (Nordgren et al.~\cite{nordgren00}), the
LD correction is relatively large ($k = \theta_{\rm LD}/\theta_{\rm UD} \simeq 1.05$),
and this could be the source of a bias at a level of a 1 to 2\% (Marengo et al.~\cite{marengo04}).
However, in the infrared, the correction is much smaller ($k \simeq 1.02$),
and the error on its absolute value is expected to be significantly below 1\%.
The majority of the Cepheid interferometric measurements
was obtained in the $H$ and $K$ bands (FLUOR/IOTA, PTI, VLTI/VINCI),
and we believe that the potential bias introduced on our fits is significantly smaller
than their stated error bars.
The final answer about the question of the limb darkening of Cepheids
will come from direct interferometric observations. The direct measurement of
the limb darkening of nearby Cepheids will soon be possible with the
AMBER instrument (Petrov et al.~\cite{petrov00}) of the VLTI.

The radial velocity data were taken from Bersier (2002). They have
been obtained with the CORAVEL spectrograph (Baranne, Mayor \&
Poncet, 1979). This instrument performs a cross-correlation of the
blue part of a star's spectrum ($3600-5200$ \AA) with the spectrum
of a red giant. A gaussian function is then fitted to the
resulting cross-correlation function, yielding the radial
velocity. 

In Paper~I, we have applied three distinct methods (orders 0, 1 and 2)
to derive the distances $d$ to seven Galactic Cepheids from
interferometric angular diameter measurements. Not all three methods
can be used to derive the distance for every star, depending on the
level of completeness and precision of the available angular diameter
measurements:
\begin{itemize}
\item{{\bf Order 0:} constant diameter model.\\
This is the most basic method, used when the pulsation of the star is
not detected. The average linear diameter $\overline{D}$ of the star
is supposed to be constant and known {\it a priori}, e.g. from a
previously published P--R relations (such as the relation derived by
GFG98).  Knowing the linear and angular radii, the only remaining
variable to fit is the distance $d$.}
\item{{\bf Order 1:} variable diameter model.\\
We still consider that the average linear diameter of the star is
known {\it a priori}, but we include in our angular diameter model the
radius variation curve derived from the integration of the radial
velocity of the star.  This method is well suited when the intrinsic
accuracy of the angular diameter measurements is too low to measure
precisely the pulsation amplitude. The distance $d$ is the only free
parameter for the fit.}
\item{{\bf Order 2:} interferometric BW method.\\
The interferometric variant of the BW method (Davis~\cite{davis79};
Sasselov et al.~\cite{sasselov94}) combines the angular amplitude of
the pulsation measured by interferometry and the linear displacement
of the stellar photosphere deduced from the integration of the radial
velocity curve to retrieve the distance of the star
geometrically. This method is also called "parallax of the pulsation".
In the fitting process, the radius curve is matched to the observed
angular diameter curve, using both the distance and linear diameter as
variables. Apart from direct trigonometric parallax, this method is
the most direct way of measuring the distance of a Cepheid.  It
requires a high precision angular diameter curve and a good phase
coverage.}
\end{itemize}

The order 0/1 methods, on one hand, and 2 on the other hand, are
fundamentally different in their assumptions, and the distance
estimates are affected by different kinds of errors. While the order 2
method errors are due to the interferometric measurement uncertainties
(mostly statistical), the order 0/1 distances carry the systematic
error bars of the assumed P--R relation. As they are fully correlated
for all stars in the sample, they cannot be averaged over the sample.
In particular, the order 0/1 diameters cannot be used to calibrate the
P--R relation, as they assume this relation to be known {\it a
priori}.

Due to its stringent requirements in terms of precision, the
interferometric BW method (order 2) was applied successfully up to now
on five Cepheids only: $\ell$\,Car (Paper~I), $\beta$\,Dor (Paper~I),
$\eta$\,Aql (Paper~I; Lane et al.~\cite{lane02}), W\,Sgr (Paper~I) and
$\zeta$\,Gem (Lane et al.~\cite{lane02}).  However, it is expected
that many more stars will be measurable with the required precision in
the near future (see Sect.~\ref{conclusion_sect}).

\section{Period--Radius relation \label{P_R_sect}}

\subsection{Method}

The Period-Radius relation (P--R) of the Cepheids takes the form of
the linear expression:
\begin{equation}
\log R = a\,\log P + b
\end{equation}
In order to calibrate this relation, we need to estimate directly the
linear radii of a set of Cepheids.  We have applied two methods to
determine the radii of the Cepheids of our sample: the interferometric
BW method, and a combination of the average angular diameter and
trigonometric parallax. While the first provides directly the average
linear radius and distance, we need to use trigonometric parallaxes to
derive the radii of the Cepheids for which the pulsation is not
detected.  We applied the {\sc Hipparcos} parallaxes (Perryman et
al.\,\cite{hip}) to all the order 0/1 measurements, except
$\delta$\,Cep, for which we considered the recent measurement by
Benedict et al.~(\cite{benedict02}).  Table~\ref{cepheid_diameters}
lists the Cepheid linear radii that we obtain.

We can use the results from both order 0/1 and 2 methods at the same
time, as the obtained linear radii obtained in this way are fully
independent from each other.  On one hand (BW method), we obtain them
considering the \emph{amplitude} of the pulsation and the radial
velocity curve, while on the other hand, they are derived from the
\emph{average} angular diameter and the trigonometric parallax.  As
the amplitude of the pulsation and the average diameter values are
distinct observables, these two methods can be used simultaneously in
the fit.

\begin{table*}
\caption{Weighted averages of the interferometric mean angular diameters
$\overline{\theta_{\rm LD}}$ and of the geometric distances $d$ to nearby
Cepheids (bold characters). These values were used to
compute the linear radii given in the last two columns.
The individual measurements used in the averaging process
are also given separately for each star.
References:
(1) Mourard et al.~(\cite{mourard97}),
(2) Nordgren et al.~(\cite{nordgren00}),
(3) Lane et al.~(\cite{lane02}),
(4) Mozurkewich et al.~(\cite{mozurkewich91}),
(5) Paper I,
(6) Benedict et al.~(\cite{benedict02}),
(7) Perryman et al.~(\cite{hip}).}
\label{cepheid_diameters}
\begin{tabular}{lccclclll}
\hline

Star & $P$\,(d) & $\log P$ & Ref.\,$\theta_{\rm LD}$ & $\overline{\theta_{\rm LD}}$\,(mas)
& Ref.\,$d$ & $d$\,(pc) & $R\,(R_\odot)$ & $\log R$ \\

\hline 
\object{$\delta$~Cep} & 5.3663 & 0.7297  & & $\bf 1.521 \pm 0.010$ & & $\bf 274^{+12}_{-11}$
& $\bf 44.8^{+1.9}_{-1.8}$ & $\bf 1.651^{+0.018}_{-0.018}$\\ 
& &  & (1) & $1.60 \pm 0.12$ \\ 
& &  & (2) & $1.52 \pm 0.01$ \\ 
& & & & & (6) & $273^{+12}_{-11}$ &  \\ 
& & & & & (7) & $301^{+64}_{-45}$ &  \\ 
\hline 
\object{X~Sgr}  & 7.0131 & 0.8459 & & $\bf 1.471 \pm 0.033$ & & $\bf 330^{+148}_{-78}$
& $\bf 52.2^{+23}_{-12}$ & $\bf 1.717^{+0.161}_{-0.118}$\\ 
& &   & (5) & $1.471 \pm 0.033$ \\ 
& & & & & (7) & $330^{+148}_{-78}$ &  \\ 
\hline 
\object{$\eta$~Aql} & 7.1768 & 0.8559 & & $\bf 1.791 \pm 0.022$ & & $\bf 308^{+27}_{-24}$
& $\bf 59.3^{+5.3}_{-4.6}$ & $\bf 1.773^{+0.037}_{-0.035}$\\ 
& &  & (2) & $1.69 \pm 0.04$ \\ 
& &  & (3) & $1.793 \pm 0.070$ & (3) & $320^{+32}_{-32}$ &  \\ 
& &  & (5) & $1.839 \pm 0.028$ & (5) & $276^{+55}_{-38}$ & \\ 
& & & & & (7) & $360^{+175}_{-89}$ &  \\ 
\hline 
\object{W~Sgr} & 7.5949 & 0.8805 & & $\bf 1.312 \pm 0.029$  & & $\bf 400^{+210}_{-114}$
& $\bf 56.4^{+30}_{-16}$ & $\bf 1.751^{+0.184}_{-0.146}$\\ 
& & & (5) & $1.312 \pm 0.029$ & (5) & $379^{+216}_{-130}$ &  \\ 
& & & & & (7) & $637^{+926}_{-237}$ &  \\ 
\hline 
\object{$\beta$~Dor} & 9.8424 & 0.9931 & & $\bf 1.884 \pm 0.024$ & & $\bf 323^{+68}_{-42}$
& $\bf 65.4^{+14}_{-8.6}$ & $\bf 1.816^{+0.083}_{-0.061}$\\ 
& &  & (5) & $1.884 \pm 0.024$ & (5) & $345^{+175}_{-80}$ &   \\ 
& & & & & (7) & $318^{+74}_{-50}$ &  \\ 
\hline
\object{$\zeta$~Gem} & 10.1501 & 1.0065 & & $\bf 1.688 \pm 0.022$ & & $\bf 362^{+37}_{-34}$
& $\bf 65.6^{+6.7}_{-6.3}$ & $\bf 1.817^{+0.042}_{-0.044}$\\ 
& &  & (2) & $1.55 \pm 0.09$ \\ 
& &  & (3) & $1.675 \pm 0.029$ & (3) & $362^{+38}_{-38}$ & \\ 
& &  & (4) & $1.73 \pm 0.05$ \\ 
& &  & (5) & $1.747 \pm 0.061$ \\ 
& & & & & (7) & $358^{+147}_{-81}$ &  \\ 
 \hline 
\object{Y~Oph} & 17.1269 & 1.2337 &  & $\bf 1.438 \pm 0.051$  & & $\bf 877^{+2100}_{-360}$
& $\bf 136^{+325}_{-56}$  & $\bf 2.132^{+0.531}_{-0.231}$\\ 
& &  & (5) & $1.438 \pm 0.051$ \\ 
& & & & & (7) & $877^{+2100}_{-360}$ &  \\ 
\hline 
\object{$\ell$~Car} & 35.5513 & 1.5509 &  & $\bf 2.988 \pm 0.012$ & & $\bf 597^{+24}_{-19}$
& $\bf 191.2^{+7.6}_{-6.0}$ & $\bf 2.281^{+0.017}_{-0.014}$\\ 
& &  & (5) & $2.988 \pm 0.012$ & (5) & $603^{+24}_{-19}$ &  \\ 
& & & & & (7) & $463^{+129}_{-83}$ &  \\ 
\hline
\end{tabular}
\end{table*}

\subsection{Calibration results}
%

\begin{figure}[t]
\centering
\includegraphics[bb=0 0 360 288, width=8.5cm]{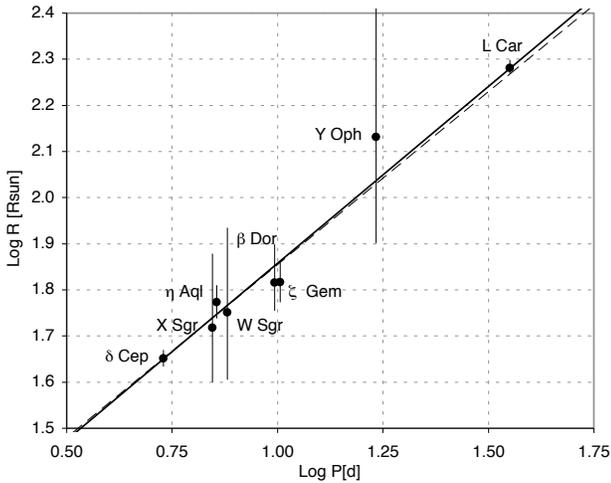}
\caption{Period-Radius diagram deduced from the interferometric
observations of Cepheids listed in Table~\ref{cepheid_diameters}.
The thin dashed line represents the
best-fit P--R relation assuming the slope of GFG98, $\log R = 0.750\
[\pm 0.024]\ \log P + 1.105\ [\pm 0.017 \pm 0.023]$.  The solid line
is the best-fit relation allowing both the slope and zero point to
vary, $\log R = 0.767\ [\pm 0.009]\ \log P + 1.091\ [\pm 0.011]$.}
\label{P_R_graph}
\end{figure}

Fig.~\ref{P_R_graph} shows the distribution of the measured diameters
on the P--R diagram, based on the values listed in
Table~\ref{cepheid_diameters}.
When we
choose to consider a constant slope of $a = 0.750 \pm 0.024$, as found
by GFG98, we derive a zero point of $b = 1.105 \pm 0.017 \pm 0.023$
(statistical and systematic errors). As a comparison, GFG98 have
obtained a value of $b = 1.075 \pm 0.007$, only -1.6\,$\sigma$ away
from our result.  The relations found by Turner \&
Burke~(\cite{turner02}) and Laney \& Stobie~(\cite{laney95}) are very
similar to GFG98, and are also compatible with our calibration within
their error bars.

\begin{table}
\caption{Period-Radius relations, assuming an expression of the form
$\log R = a\,\log P + b$. For the fitting of $b$ alone, the slope
has been assumed as known {\it a priori} from GFG98.
In this case, its error bar translates to a systematic uncertainty on
the $b$ value derived from the fit (given in brackets). References:
(1) GFG98, (2) Turner \& Burke~(\cite{turner02}),
(3) This work.}
\label{P_R_relations}
\begin{tabular}{llll}
\hline
\noalign{\smallskip}
Ref. & Fit & $a\ \pm \sigma_{\rm stat}$ & $b\ \pm \sigma_{\rm stat}\ [\pm \sigma_{\rm syst}]$ \\
\noalign{\smallskip}
\hline
\noalign{\smallskip}
(1) & & $0.750 \pm 0.024$ & $1.075 \pm 0.007$ \\
(2) & & $0.747 \pm 0.028$ & $1.071 \pm 0.025$ \\
\noalign{\smallskip}
\hline
\noalign{\smallskip}
(3) & $b$ only & & $\bf 1.105 \pm 0.017\  [\pm 0.023]$ \\
(3) & $a, b$ & $\bf 0.767 \pm 0.009$ & $\bf 1.091 \pm 0.011$ \\
\noalign{\smallskip}
\hline
\end{tabular}
\end{table}

Fitting simultaneously both the slope and the zero point to our data
set, we obtain $a = 0.767 \pm 0.009$ and $b = 1.091 \pm 0.011$.  These
values are only $\Delta a = +0.7\,\sigma$ and $\Delta b =
+1.2\,\sigma$ away from the GFG98 calibration.  Considering the
limited size of our sample, the agreement is very satisfactory.  On
the other hand, the slopes derived by Ripepi et al.~(\cite{ripepi97})
and Krockenberger, Sasselov \& Noyes~(\cite{krockenberger97}), both
around 0.60, seem to be significantly too shallow.

\section{Period--Luminosity relation \label{P_L_sect}}

\subsection{Distance estimates}

For the order 0 and 1 methods (Paper~I), we used an {\it a priori}
P--R relation (from GFG98) to predict the true linear diameter of the
Cepheids of our sample. This relation relies on the measurement of the
photometric flux, effective temperature (classical BW method) and
radial velocity. The apparent magnitude also intervenes in the
computation of the absolute magnitude, and therefore we cannot use
these distance estimates to calibrate the P--L relation without
creating a circular reference.  For this reason, we have considered
only the distances obtained using the interferometric BW method (order
2) for our P--L relation calibration, complemented by the Benedict et
al.~(\cite{benedict02}) trigonometric parallax of $\delta$\,Cep.

\subsection{Absolute magnitudes}

The average apparent magnitudes in $V$ and $K$ of $\delta$\,Cep were
computed via a Fourier series fit of the data from Moffett \&
Barnes~(\cite{moffett84}) and Barnes et al.~(\cite{barnes97}) for the
$K$ band and Barnes et al.~(\cite{barnes97}) for the $V$ band.  The
sources for the other apparent magnitudes are given in Paper~I (Table
1).  Following Fouqu\'e et al.~(\cite{fouque03}), the extinction
$A_{\lambda}$ has been computed using the relations:
\begin{equation}
A_{\lambda} = R_{\lambda}\,E_{B-V}
\end{equation}
\begin{equation}
R_{V} = 3.07 + 0.28\,(B-V)_0 + 0.04\,E_{B-V}
\end{equation}
\begin{equation}
R_{K} = R_V/11 \simeq 0.279
\end{equation}
The resulting extinction values are listed in Table~\ref{cepheid_mag_extinct}, and
the final absolute magnitudes $M_{\lambda}$ of the Cepheids of our
sample are listed in Table~\ref{cepheid_abs_mag}.

\begin{table}
\caption{Apparent magnitudes and extinctions in the
$K$ and $V$ bands for the Cepheid whose distances
have been measured directly by interferometry.
$(B-V)_0$ is the mean $(B-V)$ index
as reported in the online database by Fernie et al.~(\cite{fernie95b}).
The $E_{B-V}$ values were taken from Fernie~(\cite{fernie90}).
The extinctions  in the $K$ and $V$ bands are given respectively
in the "$A_K$" and "$A_V$" columns, in magnitudes.}
\label{cepheid_mag_extinct}
\begin{tabular}{lcccccc}
\hline
\noalign{\smallskip}
Star & $(B-V)_0$ & $E_{B-V}$ & $m_{\rm K}$ & $A_K$ & $m_{\rm V}$ &
$A_V$ \\
\hline
\noalign{\smallskip}
$\delta$\,Cep & 0.66 & 0.09 & 2.31 & 0.03 & 3.99 & 0.30 \\
\noalign{\smallskip}
$\eta$\,Aql & 0.79 & 0.15 & 1.97 & 0.04 & 3.94 & 0.49 \\
\noalign{\smallskip}
W\,Sgr & 0.75 & 0.11 & 2.82 & 0.03 & 4.70 & 0.36 \\
\noalign{\smallskip}
$\beta$\,Dor & 0.81 & 0.04 & 1.96 & 0.01 & 3.73 & 0.15 \\
\noalign{\smallskip}
$\zeta$\,Gem & 0.80 & 0.02 & 2.11 & 0.01 & 3.93 & 0.06 \\
\noalign{\smallskip}
$\ell$\,Car & 1.30 & 0.17 &1.09 & 0.05 & 3.77 & 0.58 \\
\noalign{\smallskip}
\hline
\end{tabular}
\end{table}

\begin{table}
\caption{Absolute magnitudes of Cepheids measured exclusively using
the interferometric Baade-Wesselink method, except for $\delta$\,Cep, whose
parallax was taken from Benedict et al.~(\cite{benedict02}).
The same error bars apply to the $K$ and $V$ band absolute magnitudes.
The Cepheid periods are listed in Table~\ref{cepheid_diameters}.
References:
(1) Lane et al.~(\cite{lane02}),
(2) Benedict et al.~(\cite{benedict02}),
(3) Paper I.}
\label{cepheid_abs_mag}
\begin{tabular}{lcrrccr}
\hline
\noalign{\smallskip}
Star & Ref. & $d$ & $\pm \sigma$
& $M_K$ & $M_V$ & $\pm \sigma$ \\
\noalign{\smallskip}
\hline
\noalign{\smallskip}
$\delta$\,Cep & {(2)} & $273$ & $^{+12}_{-11}$& {-4.90} & {-3.49} & $^{+0.09}_{-0.09}$ \\
\noalign{\smallskip}
$\eta$\,Aql & {(1)} & $320$ & $^{+32}_{-32}$ & {-5.60} & {-4.08} & $^{+0.23}_{-0.21}$ \\
\noalign{\smallskip}
$\eta$\,Aql & {(3)} & $276$ & $^{+55}_{-38}$ & {-5.28} & {-3.76} & $^{+0.32}_{-0.39}$ \\
\noalign{\smallskip}
W\,Sgr & {(3)} & $379$ & $^{+216}_{-130}$ & {-5.10} & {-3.56} & $^{+0.91}_{-0.98}$ \\
\noalign{\smallskip}
$\beta$\,Dor & {(3)} & $345$ & $^{+175}_{-80}$ & {-5.74} & {-4.10} & $^{+0.57}_{- 0.89}$ \\
\noalign{\smallskip}
$\zeta$\,Gem & {(1)} & $362$ & $^{+38}_{-38}$ & {-5.69} & {-3.92} & $^{+0.24}_{-0.22}$ \\
\noalign{\smallskip}
$\ell$\,Car & {(3)} & $603$ & $^{+24}_{-19}$ & {-7.86} & {-5.72} & $^{+0.07}_{-0.08}$ \\
\noalign{\smallskip}
\hline
\end{tabular}
\end{table}

\subsection{Calibration of the P--L relation \label{P_L_sect_1} }

We have considered for our fit the P--L slope measured on LMC
Cepheids. This is a reasonable assumption, as it can be measured
precisely on the Magellanic Clouds Cepheids, and in addition our
sample is currently too limited to derive both the slope and the $\log
P = 1$ reference point simultaneously.

Recently, Fouqu\'e et al.~(\cite{fouque03}) have revised the P--L
slopes derived from the large OGLE2 survey (Udalski et
al.~\cite{udalski99}), and obtain values of $\alpha_V = -2.774 \pm
0.042$ and $\alpha_K = -3.215 \pm 0.037$.  These values are consistent
within their error bars with LPG99 ($\alpha_V = -2.77 \pm 0.08$),
GFG98 ($\alpha_V = -2.769 \pm 0.073$, $\alpha_K = -3.267 \pm 0.042$)
and Sasselov et al.~(\cite{sasselov97}; $\alpha_V = -2.78 \pm 0.16$).
Considering this consensus, we have chosen to use the slope from GFG98
to keep the consistence with the P--R relation assumed in Paper~I.

\begin{figure}[t]
\centering
\includegraphics[bb=0 0 360 288, width=8.5cm]{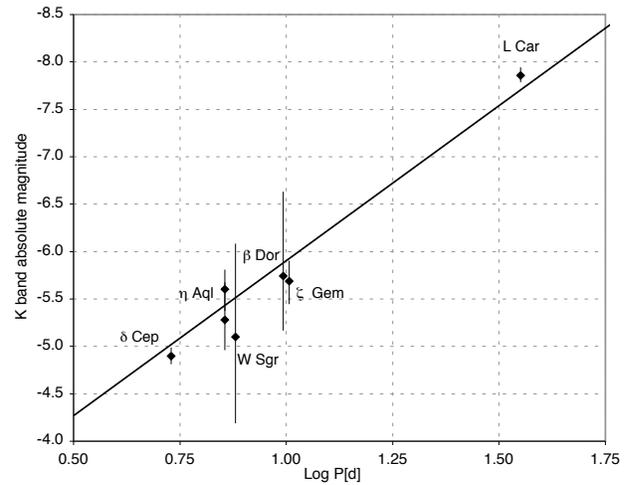}
\caption{Period-Luminosity diagram in the $K$ band using
only interferometric BW distances and the $\delta$\,Cep parallax
listed in Table~\ref{cepheid_abs_mag}. The solid line represents
the best-fit P--L relation using the slope derived by GFG98
(classical least-squares fit: the individual measurements
are weighted by the inverse of their variance).}
\label{P_L_K2}
\end{figure}

\begin{figure}[t]
\centering
\includegraphics[bb=0 0 360 288, width=8.5cm]{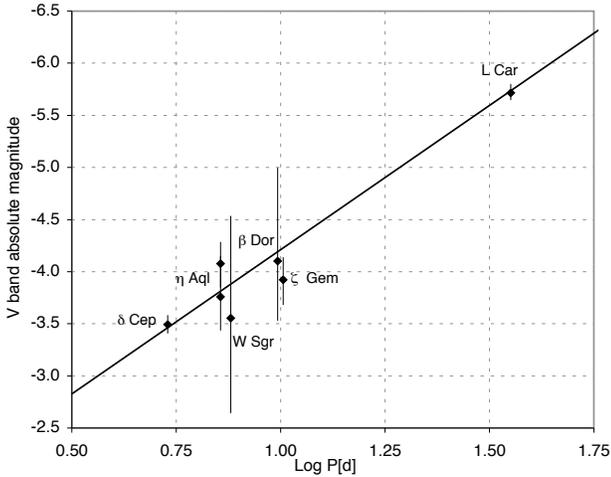}
\caption{Period-Luminosity diagram in the $V$ band (slope from GFG98).}
\label{P_L_V2}
\end{figure}

Tables~\ref{P_L_relation_K} and \ref{P_L_relation_V} report the
results of our calibrations of the P--L relations, and the positions
of the Cepheids on the P--L diagram are shown on Fig.~\ref{P_L_K2} and
Fig.~\ref{P_L_V2}.  The final $\log P = 1$ reference points are given
in bold characters in Tables~\ref{P_L_relation_K} and
\ref{P_L_relation_V}.  Our calibrations differ from GFG98 by $\Delta
b_K = +0.20$\,mag in the $K$ band, and $\Delta b_V = +0.14$\,mag in
$V$, corresponding to $+3.0$ and $+1.8\,\sigma$, respectively.  The
sample is dominated by the high precision $\ell$\,Car and
$\delta$\,Cep measurements. When these two stars are removed from the
fit, the difference with GFG98 is slightly increased, up to $+0.25$
and $+0.30$\,mag, though the distance in $\sigma$ units is reduced
($+1.3$ and $+1.5$).
From this agreement, $\ell$\,Car and $\delta$\,Cep do not
appear to be systematically different from the other Cepheids of our sample.

It is difficult to conclude firmly to a significant discrepancy
between GFG98 and our results, as our sample is currently too limited
to exclude a small-statistics bias. However, if we assume an intrinsic
dispersion of the P--L relation $\sigma_{\rm PL} \simeq 0.1$\,mag, as
suggested by GFG98, then our results point toward a slight
underestimation of the absolute magnitudes of Cepheids by these
authors.  On the other hand, we obtain precisely the same $\log P = 1$
reference point value in $V$ as Lanoix et al.~(\cite{lanoix99}, using
parallaxes from {\sc Hipparcos}).  The excellent agreement between
these two fully independent, geometrical calibrations of the P--L
relation is remarkable.

\subsection{P--L relation slopes in the Galaxy and in the LMC}

The question of the difference in slope between the Galactic and LMC
Cepheid P--L relations has recently been discussed by Fouqu\'e et
al.~(\cite{fouque03}) and Tammann et al.~(\cite{tammann03}).  These
authors conclude that the Galactic slopes are significantly steeper
than their LMC counterparts.  For example, Tamman et
al.~(\cite{tammann03}) obtain $\alpha_V [{\rm Gal}] = -3.14 \pm 0.10$,
while Fouqu\'e et al.~(\cite{fouque03}) derive $\alpha_V [{\rm Gal}] =
-3.06 \pm 0.11$ and $\alpha_V [{\rm LMC}] = -2.774 \pm 0.042$.

However, our fit is largely insensitive to the precise value assumed
for the P--L relation slope. Considering the steeper Tammann et
al.~(\cite{tammann03}) slope, we obtain a best fit $\log P = 1$
absolute magnitude of $\beta_V = -4.211 \pm 0.075 \pm 0.001$,
identical to the calibration obtained using the GFG98 slope.  The
small systematic error bar that we obtain on $\beta_V$ (corresponding
to the $\pm 0.10$ error on $\alpha_V$) shows the weakness of the
correlation between $\alpha$ and $\beta$ in our fit.  However, the
reduced $\chi^2$ of the fit is significantly larger with this steeper
slope ($\chi^2_{\rm red} = 1.25$) than with the LMC slope from GFG98
($\chi^2_{\rm red} = 0.53$).

\begin{table}
\caption{Period-Luminosity relation intercept $\beta_K$
for a 10\,days period Cepheid ($\log P = 1$), in the $K$ band.
We assume an expression of  the form $M_{K} = \alpha_K\,(\log P-1) + \beta_K$.
The slope value is taken from GFG98 ($\alpha_K = -3.267 \pm 0.042$).
The systematic error is  given in brackets, and corresponds to
the uncertainty on the GFG98 slope.}
\label{P_L_relation_K}
\begin{tabular}{lccc}
\hline
\noalign{\smallskip}
Ref. & $\beta_K$ & $\pm \sigma_{\rm stat}$ & $\pm \sigma_{\rm syst}$ \\
\noalign{\smallskip}
\hline
\noalign{\smallskip}
GFG98 & $-5.701$ & $\pm 0.025$ \\
\noalign{\smallskip}
\hline
\noalign{\smallskip}
This work, all stars & $\bf -5.904$ & $\bf \pm 0.063$ & $\bf \pm 0.005$ \\
Without $\delta$\,Cep and $\ell$\,Car & $-5.956$ & $\pm 0.191$ & $\pm 0.006$ \\
\noalign{\smallskip}
\hline
\end{tabular}
\end{table}

\begin{table}
\caption{Period-Luminosity relation intercept $\beta_V$ ($\log P = 1$)
in the $V$ band, derived using the GFG98 slope ($\alpha_V = -2.769 \pm 0.073$).}
\label{P_L_relation_V}
\begin{tabular}{lccc}
\hline
\noalign{\smallskip}
 & $\beta_V$ & $\pm \sigma_{\rm stat}$ & $\pm \sigma_{\rm syst}$\\
\noalign{\smallskip}
\hline
\noalign{\smallskip}
GFG98 & $-4.063$ & $\pm 0.034$ & \\
LPG99 & $-4.21$ & $\pm 0.05$ & \\
\noalign{\smallskip}
\hline
\noalign{\smallskip}
This work, all stars & $\bf -4.209$ & $\bf \pm 0.075$ & $\bf \pm 0.001$ \\
Without $\delta$\,Cep and $\ell$\,Car & $-4.358$ & $\pm 0.197$ & $\pm 0.010$ \\
\noalign{\smallskip}
\hline
\end{tabular}
\end{table}

\subsection{The distance to the LMC \label{lmc_dist}}

The apparent magnitudes in $V$ and $K$ of a 10\,days period Cepheid in
the Large Magellanic Cloud (LMC) derived by Fouqu\'e et
al.~(\cite{fouque03}) from the OGLE Cepheids are ${\rm ZP}_K = 12.806
\pm 0.026$ and ${\rm ZP}_V = 14.453 \pm 0.029$. These authors assumed
in their computation a constant reddening of $E(B-V) = 0.10$ for all
the LMC Cepheids they have used (more than 600).
Our calibrations of the Galactic Cepheids P--L relations in $K$ and $V$
thus implies LMC distance moduli of
$\mu_K = 18.71 \pm 0.07$ and $\mu_V = 18.66 \pm 0.08$, respectively.

From a large number of photometric measurements of LMC and SMC
Cepheids obtained in the framework of the EROS programme, Sasselov et
al.~(\cite{sasselov97}) have shown that a $\delta \mu$ correction has
to be applied to the LMC distance modulus to account for the
difference in metallicity between the LMC and the Galactic
Cepheids. They have determined empirically a value of:
\begin{equation}
\delta \mu = \mu_{\rm true} - \mu_{\rm observed} = -0.14 \pm 0.06
\end{equation}
We would like to point out that this correction has been questioned
by Udalski et al.~(\cite{udalski01}), based on Cepheid observations in
a low metallicity galaxy (IC\,1613), and its amplitude is still under
discussion (Fouqu\'e et al.~\cite{fouque03}).

Averaging our $K$ and $V$ band zero point values (without reducing the
uncertainty, that is systematic in nature), we obtain a final LMC
distance modulus of
$\mu_0 = 18.55 \pm 0.10$.
This value is only $+0.8\,\sigma$ away from the $\mu_0 = 18.46 \pm
0.06$ value obtained by GFG98, and $-1\,\sigma$ from the $\mu_0 =
18.70 \pm 0.10$ value derived of Feast \& Catchpole~(\cite{feast97}).
It is statistically identical to the LMC distance used by Freedman et
al.\,(\cite{freedman01}) for the {\it HST Key Project}, $\mu_0 = 18.50
\pm 0.10$.  Alternatively, if we consider the smaller metallicity
correction of $\delta \mu = 0.06 \pm 0.06$ proposed by GFG98, we
obtain a distance modulus of $\mu_0 = 18.63 \pm 0.10$.

\section{Conclusion and perspectives \label{conclusion_sect}}

We have confirmed in this paper the P--R relation of GFG98 and Turner
\& Burke~(\cite{turner02}), to a precision of $\Delta(\log R) = \pm
0.02$.  We also derived an original calibration of the P--L relations
in $K$ and $V$, assuming the slopes from GFG98 that were established
using LMC Cepheids.  Our P--L relation calibration yields a distance
modulus of $\mu_0 = 18.55 \pm 0.10$ for the LMC, that is statistically
identical to the value used by Freedman et al.~(\cite{freedman01}) for
the {\it HST Key Project}. We would like to emphasize that this
result, though encouraging, is based on six stars only (seven
measurements, dominated by two stars), and our sample needs to be
extended in order to exclude a small-number statistics bias.  In this
sense, the P--L calibration presented here should be considered as an
intermediate step toward a final and robust determination of this
important relation by interferometry.


While our results are very encouraging, the calibration of
the PR and PL relations as described here may still be affected by
small systematic errors. In particular the method relies on the
fact that the displacements measured through interferometry and
through spectroscopy (integration of the radial velocity curve)
are in different units (milli-arcseconds and kilometers
respectively) but are the same physical quantity. This may not be
the case. The regions of a Cepheid's atmosphere where the lines
are formed do not necessarily move homologously with the region
where the $K$-band continuum is formed. This means that the two
diameter curves may not have exactly the same amplitude; there
could even be a phase shift between them.
As discussed in Sect.~\ref{sect_context}, the limb darkening
could also play a role at a level of $\simeq 1\%$.
A full exploration of these effects is far beyond the scope of this paper.
We can nevertheless put an upper bound on the systematic error that could
result from this mismatch. Our PL relation can be compared to that
derived from Cepheids in open clusters, whose distances are
obtained via main sequence fitting. The two distance scales are in
excellent agreement (Gieren \& Fouqu\'e 1993; Turner \& Burke,
2002). These distances are consistent with a Pleiades distance
modulus of 5.56; if anything they are slightly larger.


The availability of 1.8\,m Auxiliary Telescopes (Koehler et
al.~\cite{koehler02}) on the VLTI platform in 2004, to replace the
current 0.35\,m Test Siderostats, will allow to observe many Cepheids
with a precision at least as good as the observations of $\ell$\,Car
reported in Paper~I (angular diameters accurate to 1\%). In addition,
the AMBER instrument (Petrov et al.~\cite{petrov00}) will extend the
VLTI capabilities toward shorter wavelengths ($J$ and $H$ bands),
thus providing higher spatial resolution than VINCI ($K$ band).  The
combination of these two improvements will extend significantly the
accessible sample of Cepheids, and we expect that the distances to
more than 30 Cepheids will be measurable with a precision better than
$\pm 5$\,\%.
This will provide a high precision calibration of both the
$\log P = 1$ reference point (down to $\pm
0.01$\,mag) and the slope of the Galactic Cepheid P--L.
As the galaxies hosting the Cepheids used in the {\it Key
Project} are close to solar metallicity in average
(Feast~\cite{feast01}), this Galactic calibration will allow to bypass
the LMC step in the extragalactic distance scale. Its attached
uncertainty of $\pm 0.06$ due to the metallicity correction of the LMC
Cepheids will therefore become irrelevant for the measurement of
$H_0$.

\begin{acknowledgements}
DB acknowledges partial support from NSF grant AST-9979812.  PK
acknowledges support from the European Southern Observatory through a
post-doctoral fellowship.  Based on observations collected at the VLT
Interferometer, Cerro Paranal, Chile, in the framework of the ESO
shared-risk programme 071.D-0425 and an unreferenced programme in P70.
The VINCI/VLTI public commissioning data reported in this paper have
been retrieved from the ESO/ST-ECF Archive (Garching, Germany).
This research has made use of the SIMBAD database at CDS, Strasbourg (France).
We are grateful to the ESO VLTI team, without whose efforts no observation
would have been possible.
\end{acknowledgements}

{}
\end{document}